\documentclass[twocolumn,usenatbib]{aa}
\usepackage{graphicx}
\usepackage{epsfig,graphicx,rotating,portland,fancyheadings,longtable,natbib}
\usepackage{txfonts}

\def\ls{{_<\atop^{\sim}}}
\def\gs{{_>\atop^{\sim}}}

\begin{document}
   \title{UVES/VLT high resolution spectroscopy of GRB 050730
afterglow: probing the features of the GRB environment \thanks{Based
on observations collected at the European Southern Observatory, ESO,
the VLT/Kueyen telescope, Paranal, Chile, in the framework of programs
075.A-0603}}

\author{V. D'Elia$^1$, F. Fiore$^1$, E.J.A. Meurs$^2$, G. Chincarini$^{3,4}$,
A. Melandri$^5$, L. Norci$^6$, L. Pellizza$^7$, R. Perna$^{8}$, S. Piranomonte$^1$,
L. Sbordone $^1$, L. Stella$^1$, G. Tagliaferri $^3$, S.D. Vergani$^{2,6}$, P. Ward$^2$,
L. Angelini$^9$, L. A. Antonelli$^1$, D.N. Burrows$^{10}$,
S. Campana$^3$, M. Capalbi$^{11}$, A. Cimatti$^{12}$, E. Costa$^{13}$,
G. Cusumano$^{14}$, M. Della Valle$^{12}$, P. Filliatre$^7$, A. Fontana$^1$
F. Frontera$^{15}$, D. Fugazza$^3$, N. Gehrels$^9$, T. Giannini$^1$,
P. Giommi$^{11}$, P. Goldoni$^{16}$, D. Guetta$^1$, G. Israel$^1$,
D. Lazzati$^{8}$, D. Malesani$^{17}$, G. Marconi$^{18}$, K. Mason$^{19}$, 
S. Mereghetti$^{20}$, F. Mirabel, $^{21}$
E. Molinari$^3$, A. Moretti$^3$, J. Nousek$^{10}$, M. Perri$^{11}$, 
L. Piro$^{13}$, G. Stratta$^{11}$, V. Testa$^1$, M. Vietri$^{22}$
}

\institute
{$^1$INAF-Osservatorio Astronomico di Roma, Via Frascati 33, I-00044 Monteporzio Catone, Italy;\\ 
$^2$ Dunsink Observatory, Castleknock, Dublin 15, Ireland;\\ 
$^3$ INAF, Osservatorio Astronomico di Brera, via E. Bianchi 46, 23807 Merate (LC), Italy;\\ 
$^4$ Universita' di Milano Bicocca, Piazza della Scienza 3, 20126 Milano, Italy.\\
$^5$ Astronomic Research Institute, Liverpool John Moores University, Twelve Quays House, Egerton Wharf, Birkenhead, CH41 1LD, UK; \\ 
$^6$ School of Physical Sciences and NCPST, Dublin City University, Glasnevin, Dublin 9, Ireland\\
$^7$ CEA Saclay, DSM/DAPNIA/Service d'Astrophysique, Bat. 709, L'Orme des Merisiers, 91191 Gif-sur-Yvette, Cedex, France; \\ 
$^8$ JILA, Campus Box 440, University of Colorado, Boulder, CO 80309-0440, USA; \\
$^9$ NASA-Goddard Space Flight Center, Greenbelt, Maryland, 20771, USA; \\
$^{10}$ Department of Astronomy and Astrophysics, Pennsylvania State University, University Park, Pennsylvania 16802, USA; \\
$^{11}$ ASI Science Data Center, Via G. Galilei, I-00044, Frascati(Roma), Italy; \\
$^{12}$ INAF-Osservatorio Astronomico di Arcetri, Largo E. Fermi 5, 50125, Firenze, Italy; \\
$^{13}$ INAF-Roma, Via Fosso del Cavaliere 100, I-00133, Roma, Italy; \\
$^{14}$ INAF-Istituto di Astrofisica Spaziale e Fisica Cosmica di Palermo, Via U. La Malfa 153, I-90146, Palermo, Italy; \\
$^{15}$ Universita' di Ferrara, Via Paradiso 12, 44100, Ferrara, Italy; \\
$^{16}$ UMR 7164, 11 Place M. Berthelot, 75231, Paris, France; \\
$^{17}$ International School for Advanced Physics, Via Beirut 2-4, I-34014, Trieste, Italy; \\
$^{18}$ European Southern Observatory, Casilla 19001, Santiago, Chile; \\
$^{19}$ PPARC, Polaris House, North Star Avenue, Swindon SN2 15Z, UK; \\
$^{20}$ INAF-Istituto di Astrofisica Spaziale e Fisica Cosmica, Via Bassini, 15, I-20133, Milano, Italy; \\
$^{21}$ European Southern Observatory-Vitacura, Casilla 19001, Santiago 19, Chile;\\
$^{22}$ Scuola Normale Superiore, Piazza dei Cavalieri 7, 56126, Pisa, Italy;\\
}

  \abstract 
  {} 
  {The aim of this
  paper is to study the Gamma Ray Burst (GRB) environment through the
  analysis of the optical absorption features due to the gas
  surrounding the GRB.  
  }  
  {To this purpose we
  analyze high resolution spectroscopic observations (R=20000-45000,
  corresponding to 14 km/s at 4200\AA\ and 6.6 km/s at 9000\AA) of the
  optical afterglow of GRB050730, obtained with UVES@VLT $\sim 4$
  hours after the GRB trigger.}  
  {The spectrum shows that the ISM of the GRB host galaxy at z = 3.967 is
  complex, with at least five components contributing to the main
  absorption system. We detect strong {\ion{C}{II*}}, {\ion{Si}{II*}},
  {\ion{O}{I*}} and {\ion{Fe}{II*}} fine structure absorption lines
  associated to the second and third component.}  
  {For the first three
  components we derive information on the relative distance from the
  site of the GRB explosion.  Component 1, which has the longest
  wavelength, highest positive velocity shift, does not present any
  fine structure nor low ionization lines; it only shows very high
  ionization features, such as {\ion{C}{IV}} and {\ion{O}{VI}},
  suggesting that this component is very close to the GRB site.  From
  the analysis of low and high ionization lines and fine structure
  lines, we find evidences that the distance of component 2 from the
  site of the GRB explosion is 10-100 times smaller than that of
  component 3.  We evaluated the mean metallicity of the z=3.967
  system obtaining values $\approx 10^{-2}$ of the solar metallicity
  or less. However, this should not be taken as representative of the
  circumburst medium, since the main contribution to the hydrogen
  column density comes from the outer regions of the galaxy while that
  of the other elements presumably comes from the ISM closer to the
  GRB site. Furthermore, difficulties in evaluating dust depletion
  correction can modify significantly these values.

  The mean [C/Fe] ratio agrees well with that expected by single
  star-formation event models. Interestingly the [C/Fe] of component 2 is
  smaller than that of component 3, in agreement with GRB dust
  destruction scenarios, if component 2 is closer than component 3 to
  the GRB site.}

\authorrunning {D'Elia et al.}
\titlerunning {UVES/VLT high resolution spectroscopy of GRB050730}

\maketitle
%

\section{Introduction}

For a few hours after their onset, Gamma Ray Bursts (GRBs) are the
brightest beacons in the far Universe, offering a superb opportunity
to investigate both GRB physics and high redshift galaxies.  Tens of
minutes after a GRB, its optical afterglow can be as bright as
magnitude 13--15.  High resolution (a few tens of km/s in the optical
band), high quality (signal to noise $>10$ per resolution element)
spectra can therefore be gathered, provided that the afterglow is
observed on such a short time scale by 8m class telescopes.

Fiore et al. (2005) studied the UVES high resolution spectroscopy of
GRB020813 and GRB021004 afterglows; they showed that the InterStellar
Medium (ISM) of GRB host galaxies is complex when resolved down to a
width of a few tens of km/s, with many components contributing to each
main absorption system, and spanning a total velocity range of up to
thousands of km/s. Velocity ranges of hundreds of km/s were found
on the high resolution spectra obtained with MIKE@MagellanII and 
HIRES@Keck on GRB051111 and GRB050730 
(Chen et al. 2005, Prochaska, Chen \& Bloom 2006, PCB06 hereafter,
Penprase et al. 2006).

The absorption systems can be divided into three broad
categories. First, those associated with the GRB surrounding medium,
second, those produced by the ISM of the host galaxy along the line
of sight but distant enough not to be strongly modified by the 
GRB, and third the intergalactic matter along the line of sight. 
We discuss these three systems in turn.

The physical, dynamical and chemical status of the circumburst medium
in the star-forming region hosting the GRB progenitor can be modified
by the explosive event, through shock waves and ionizing
photons. Blueshifted absorbers may be ionized and even radiatively
and/or collisionally accelerated by the GRB and its afterglow.
Circumstellar absorption lines in gamma-ray burst afterglow spectra
can be used to determine the main properties of the star progenitors
(van Marle, Langer,\& Garcia-Segura 2005). The variability of the
intensity of absorption lines can be used to determine the location
and density of the absorber, following Perna \& Loeb (1998),
Boettcher, Fryer \& Dermer (2002), Draine \& Hao (2002), Mirabal et
al. (2002), Perna \& Lazzati (2002), Perna, Lazzati \& Fiore
(2003). The temperature and density of the absorbing gas can be also
accurately determined by studying particular line ratios, namely those
of a fine structure transition to its ground state.

High resolution spectra do not reveal only the circumburst
matter. They can also be used to probe the ISM of the host galaxy
along the line of sight.  The study of the ISM of z$\gs1$ galaxies has
so far relied upon Lyman-break galaxies (LBGs) at z$=3-4$ (see
e.g. Steidel et al.  1999) and galaxies which happen to be along lines
of sight to bright background quasars (QSOs).  However, LBGs are
characterized by pronounced star-formation and their inferred chemical
abundances may be related to these regions rather than being
representative of typical high-z galaxies.  Faint metal line systems
along the line of sight to quasars probe mainly galaxy haloes, rather
than their bulges or discs. Taking advantage of ultra-deep {\it
Gemini} multi-object spectrograph observations, Savaglio et al. (2004,
2005) studied the ISM of a sample of faint K band selected galaxies at
1.4$<$z$<$2.0, finding MgII and FeII abundances much higher than in
QSO systems and similar to those in GRB host galaxies. Such studies
will hardly be extended to higher redshift with the present generation
of 8m class telescopes, because of the faintness of high-z
galaxies. On the other hand, GRB afterglows provide an independent
tool to study the ISM of high z galaxies.

Finally, GRB afterglow high resolution spectra can be used to probe
the Ly-$\alpha$ forest and the high-z intergalactic medium. Using GRBs as
remote beacons opens up the opportunity to highlight any deviation
from what is already known from quasar forests.  For example the 
so-called ``proximity effect'' (the feedback of QSOs on their local IGM)
should be much reduced for GRBs.

The need to disentangle the contribution to the absorption spectra
coming from these three different categories renders high resolution
spectroscopy the ideal technique to study GRB 
optical afterglows.

Therefore we set up a program to observe bright optical
afterglows of promptly localized GRBs with UVES@VLT, taking 
advantage of the capabilities of the GRB-dedicated
{\it Swift} satellite. {\it Swift} was launched in November 2004, and
provides GRB positions ($2-3\arcmin$ precision) in $<10$~s, X--ray
afterglow positions ($5\arcsec$ precision) in $<100$~s, and an optical
finding chart in $<300$~s (few arcsec to sub-arcsec positional 
accuracy). These characteristics, together with the possibility of 
quick follow-up with UVES@VLT in RRM (Rapid Response Mode), 
offers the opportunity to obtain high resolution spectra of GRBs 
with delays from a few hours down to a few tens of minutes from the 
event.  

To date, five GRBs have been observed with UVES@VLT after a {\it Swift} 
trigger: GRB050730, GRB050820A, GRB050922C, GRB060418 and GRB060607A. 

In this paper, we present high resolution spectroscopy of GRB050730.
This GRB was detected by {\it Swift} on 2005 July 30 19:58:23 UT
(Holland et al. 2005, Sota et al. 2005) at RA(J2000) $=14^h08^m17^s.14$ 
and DEC(J2000) $= -03^d46^{'}17^{''}.8$. $133$ seconds later, 
{\it Swift}'s XRT started to observe the X-ray afterglow in WT mode
(Barthelmy et al. 2005). Chen et al. (2005) found the redshift 
of the GRB at $z =3.967$ with the MIKE echelle spectrograph on Magellan II,
starting the observation $\sim 4$ hours after the burst.  The redshift
was confirmed by Rol et al. (2005) using the ISIS spectrograph on the
William Herschel Telescope at the Observatorio del Roque de los
Muchachos on La Palma.

Starling et al. (2005) show an early WHT ISIS optical spectroscopy of
GRB050730, finding that the host is a low metallicity galaxy, with low
dust content. PCB06 present an analysis of the MIKE high resolution
spectrum of GRB050730, treating the various absorption features as
single Voigt profiles. Our main goal is instead that of trying to
disentangle the relative contribution of the various components (in
velocity space) contributing to each absorption system, making a step
forward with respect to Chen, et al. (2005) and PCB06.

We focus here on the study of the local medium surrounding the GRB, 
and in particular on the analysis of the fine structure absorption lines
produced by the gas in the host galaxy. The comparison of such 
absorption features and their corresponding ground states yields 
information about the gas density, the temperature and the 
radiation field.  A detailed analysis of the intervening absorption systems 
and of the Ly-$\alpha$ forest will be presented in a forthcoming paper.

The paper is organized as follows. Section $2$ presents the
observations and data reduction; Section $3$ gives a brief description
of the absorbing systems identified from the spectra; Section $4$ is devoted to the 
analysis of the fine structure lines and to the estimate of the physical 
quantities characterizing the host galaxy gas; in Section $5$ 
the results are discussed and conclusions are drawn. 

\begin{figure}
\centering
\includegraphics[angle=-90,width=9cm]{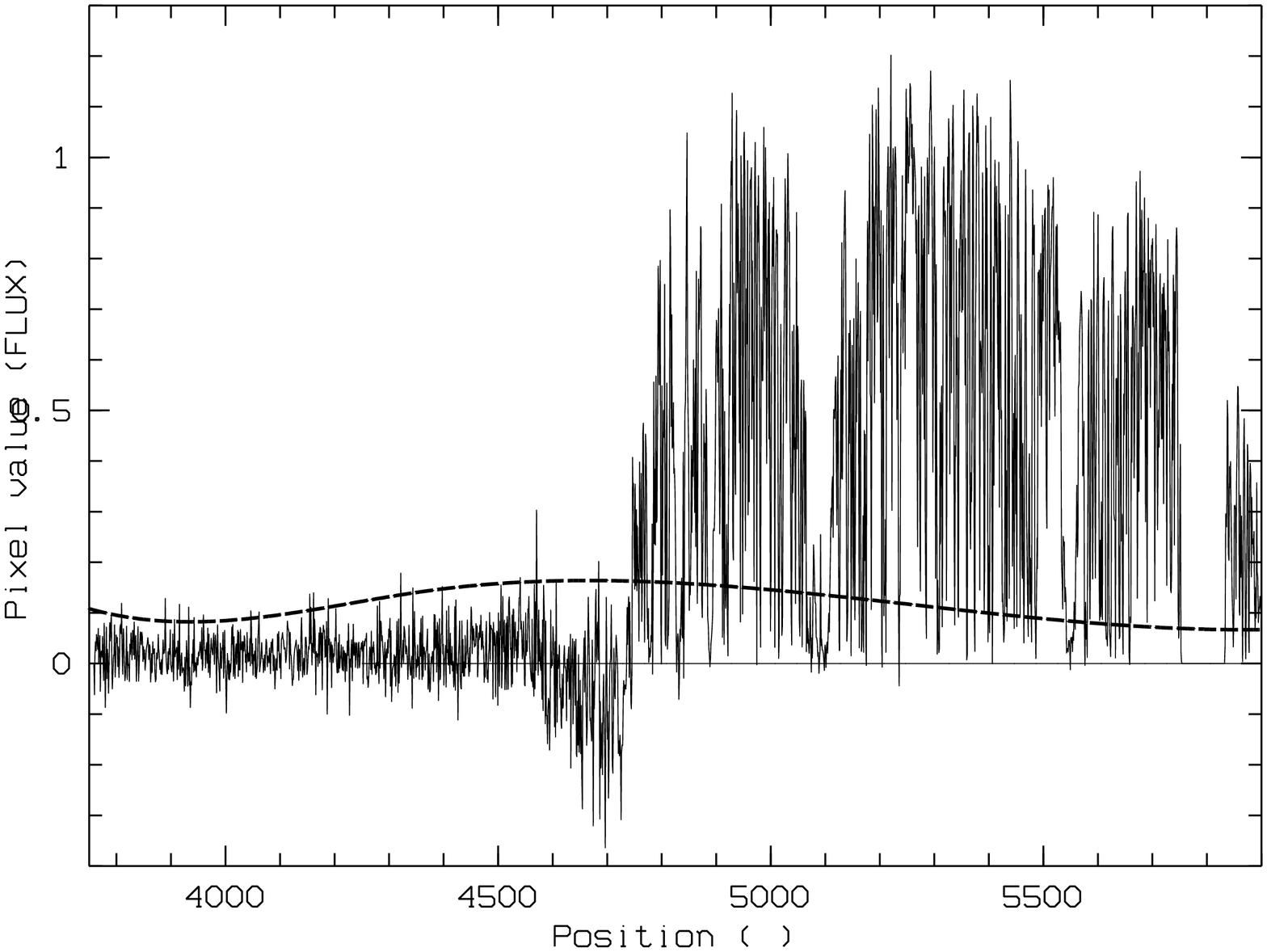}
\includegraphics[angle=-90,width=9cm]{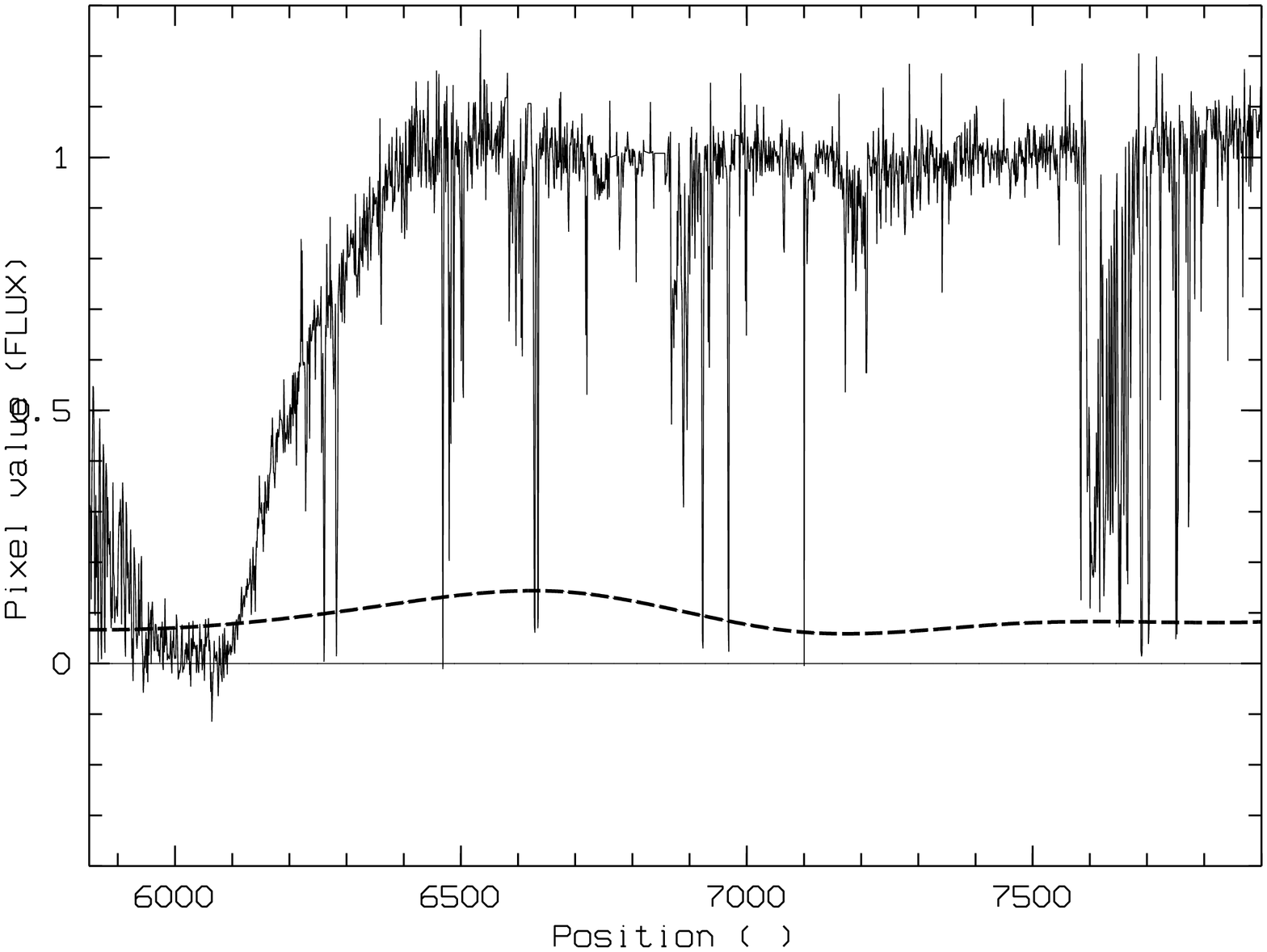}
\includegraphics[angle=-90,width=9cm]{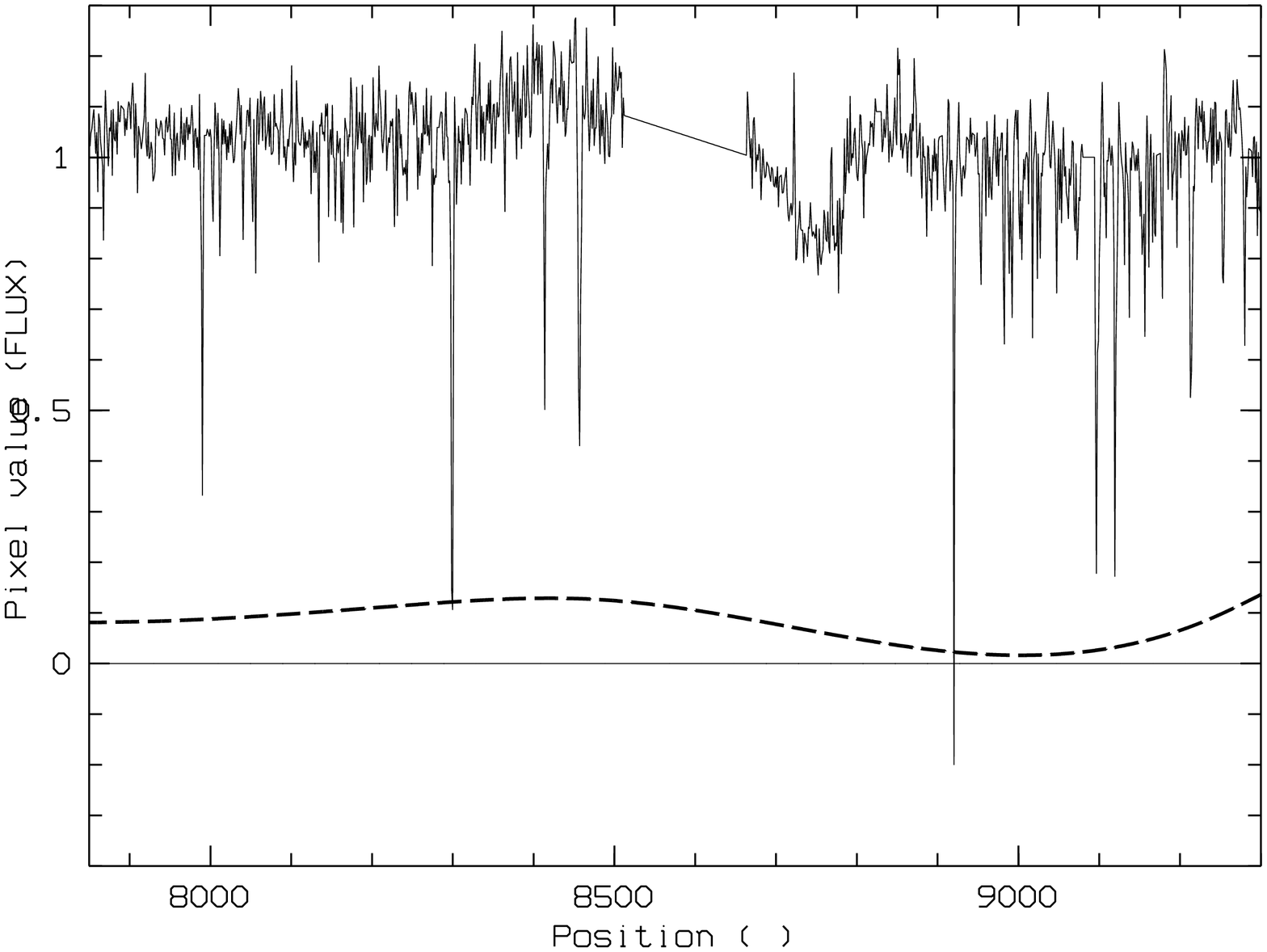}
\caption{The full high resolution spectrum of GRB050730. Dashed lines
represent the noise spectrum. The spectrum is cut-off at the bluemost
and redmost end, where the signal-to-noise ratio becomes too small.
}
\label{spe1}
\end{figure}

\section{Observations and data reduction}

In the framework of ESO programs 075.A-0603 we observed the afterglow
of GRB050730 with the high resolution UV-visual echelle spectrograph
(UVES, Dekker et al. 2000), mounted at the VLT-UT2 telescope.  Table 1
gives the log of the observations.  Both UVES dichroics were used,
employing the red as well as the blue arm.  This allowed us to achieve
a particularly wide spectral coverage, extending from $\sim$3200\AA\
to $\sim$9500\AA.  In order to maximize the signal to noise ratio
the CCD was rebinned by $2\times2$ pixels. The data reduction has been
performed using the UVES pipeline (Ballester et al. 2000). The final
useful spectra extend from about 3750\AA\ to about 9780\AA\ and were
rebinned to 0.1\AA. The resolution element, set to two pixels,
ranges then from 14 km/s at 4200\AA\ to 6.6 km/s at 9000\AA.  The
noise spectrum, used to determine the errors on the best fit line
parameters, has been calculated from the real, background subtracted
and rebinned spectrum, using line free regions. This therefore takes
into account both statistical errors and systematic errors in the
pipeline processing as well as the background subtraction.

\begin{table*}[ht]
\caption{\bf Journal of observations}
\footnotesize
\begin{tabular}{lccccccc}
\hline
Date UT  & Dichroic & B. Arm C.W. & R. Arm C.W. & slit$^{a}$ 
& seeing$^{a}$ & exposure$^{b}$ & time since GRB$^{c}$\\
\hline
07/31/05 00:07:17 & 1	& 3460\AA\ & 5800\AA\ & 1 & $\ls1$ & 50 & 4.15 \\
07/31/05 01:02:09 & 2	& 4370\AA\ & 8600\AA\ & 1 & $\ls1$ & 50 & 5.07 \\
\hline
\end{tabular}
\normalsize

$^{a}$arcsec;
$^{b}$min;
$^{c}$hr
\end{table*}

\section{Absorbing systems}

Fig. 1 shows the high resolution spectrum of GRB050730, with a cut-off
at both the bluemost and redmost end, where the signal-to-noise ratio
becomes too small. The spectrum exhibits a large number of absorption
features, belonging to different absorption systems. The main system
is due to the gas in the host galaxy, at redshift
$z_{GRB}=3.96764$. This value confirms the previous redshift
determinations for this GRBs. We identified four more intervening
absorbers at $z<z_{GRB}$ not related to the GRB host galaxy but
rather to galaxies along the line of sight.  Each system is briefly
described in the next subsections.

\subsection{The host environment at $z = 3.967$}

This is the system with the greatest number of features on the GRB line
of sight.  Fig. 1 clearly shows the Ly-$\alpha$ absorption at $\sim 6000$ \AA. 
The redshift inferred from this feature is confirmed by a large number 
of metal-line transitions. The precise value, which we calculated from 
the analysis of the narrowest lines associated with the host 
galaxy, is $z=3.96764$. The spectrum exhibits 
a large number of 
absorption lines, from neutral hydrogen (Ly-$\alpha$, Ly-$\beta$, 
Ly-$\gamma$), via neutral metal-absorption lines ({\ion{O}{I}}) 
and low ionization lines ({\ion{C}{II}}, {\ion{Si}{II}},
{\ion{Al}{II}}, {\ion{Ni}{II}}, {\ion{Fe}{II}},
{\ion{P}{II}}, {\ion{S}{II}})  to high-ionization
absorption features ({\ion{C}{IV}},
{\ion{Si}{IV}},{\ion{N}{V}}, {\ion{O}{VI}}). In
addition, strong fine structure lines ({\ion{C}{II*}},
{\ion{Si}{II*}}, {\ion{O}{I*}} {\ion{O}{II**}},
{\ion{Fe}{II*}}, {\ion{Fe}{II**}},
{\ion{Fe}{II***}}) have been identified, meaning either that
the GRB environment is composed of a high density gas or that an
intense radiation field is exciting such features.

The ISM of the host galaxy is complex, with
many components contributing to the absorption system. This complex
environment, together with a detailed study of the fine structure
lines aimed at estimating the physical quantities of the circumburst
gas, will be described in the next section.

\subsection{The intervening absorber at $z = 3.564$}

This system presents neutral hydrogen features (Ly-$\alpha$,
Ly-$\beta$), neutral metal-absorption lines ({\ion{C}{I}},
{\ion{Si}{I}}), low ionization lines ({\ion{Al}{II}},
{\ion{Mg}{II}}, {\ion{Si}{II}}, {\ion{Fe}{II}}) and
high ionization lines ({\ion{C}{IV}}, {\ion{Si}{IV}},
{\ion{N}{V}}).  The redshift calculated from these 
features is $z \sim 3.56395$.

\subsection{The intervening absorber at $z = 2.262$}

This system presents only a few neutral metal-absorption lines (OI,
SiI, AlI) and some low ionization transitions ({\ion{Al}{II}},
{\ion{Fe}{II}}). The redshift of this system is $z \sim
2.26181$.
  
\subsection{The intervening absorber at $z = 2.253$}

This is actually a double absorption system, the redshifts of the two
components being $z \sim 2.2526$ and $z \sim 2.2536$. It shows neutral
metal-absorption lines ({\ion{Si}{I}}, {\ion{Mg}{I}}),
low ionization transitions ({\ion{Si}{II}},
{\ion{Fe}{II}}, {\ion{Mg}{II}}) and possibly a
{\ion{C}{IV}} high ionization line.

\subsection{The intervening absorber at $z = 1.772$}

This is another double absorption system, the redshifts of the two
components being $z \sim 1.7723$ and $z \sim 1.7729$. The system shows
neutral metal-absorption lines ({\ion{Mg}{I}},
{\ion{Fe}{I}}) and low ionization transitions
({\ion{Mg}{II}}, {\ion{Fe}{II}}).

\section{The main system at z=3.967: physical properties of the GRB environment}

This section, as well as the rest of the paper, is devoted to the
study of the main absorption feature of the spectrum, produced by the
gas in the host environment, located close to the GRB explosion
site.

The analysis of the GRB environment is intricate due to the complexity
of the absorption lines of the spectrum, which in several cases can
hardly be fitted with a single Voigt profile. This means that many
components contribute to the gas in the GRB environment.  In other
words, several shells of gas which may be close to or far from each
other, appear mixed together in the spectrum (in velocity space). Our
main goal is to disentangle their relative contribution.

The identification of the components of the absorption lines and the
fitting procedures will be discussed in the next subsection.

\subsection{Absorption features of the GRB environment}

The gas surrounding the GRB constitutes a complex environment, and this 
is reflected in the complexity of the absorption features. The presence 
of several components is indicative of clumpy gas in the GRB
environment, composed of different absorbing regions each with
different physical properties. For example, a first component is present 
for the high ionization lines {\ion{O}{VI}} and 
(less intensely) {\ion{C}{IV}}, but not 
for {\ion{Si}{IV}} and lower ionization lines (Fig. 2). This means
that this particular component probably comes from a region very close
to the GRB site, where it experienced a very intense flux of radiation
ionizing the atoms several times.

If the absorption features are treated as a single Voigt profile, we
obtain very bad fitting parameters and very high $\chi^2$
values.  The identification of the different components is somewhat
subjective, the true message being that the geometry and kinematics of
the ISM clouds probed by the GRB line of sight are complex.  We
decided to use the {\ion{C}{IV} $\lambda$1548,1550} \AA\ and
{\ion{Si}{IV} $\lambda$1393,1402} \AA\ doublets,
which have the wider velocity range,  to identify the different components
constituting the circumburst matter. Fig. 2 shows {\ion{C}{IV}
$\lambda$1550} \AA\ at $z = 3.9676$. For this line we obtain a
good fit ($\chi^2 = 1.04$) if we assume that five different
components contribute to the absorption features.

\begin{figure}
\centering
\includegraphics[angle=-90,width=9cm]{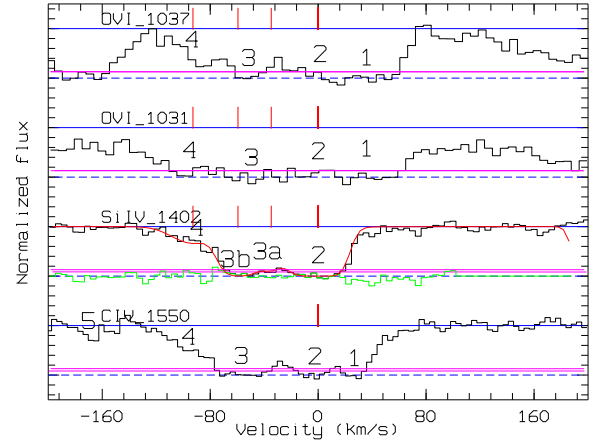}
\caption{The {\ion{C}{IV} $\lambda$1550} \AA\ and
{\ion{Si}{IV} $\lambda$1402} \AA\ absorption features together
with the {\ion{O}{VI} $\lambda$1031,1037} \AA\ doublet. Each
component is identified with progressive numbers from the highest
wavelength to with the lowest one.  }
\label{spe1}
\end{figure}

In the following, the components are numbered so that the first
component is the one with the longest wavelength and the
highest, positive, velocity shift with respect to a zero point
arbitrarily placed at z=3.96764, and so on.  We next fixed the
redshifts of the {\ion{C}{IV}} components and tried to fit the
{\ion{Si}{IV}} doublet with the same five Voigt profiles. We did not
obtain a good fit for the following reasons. First, the {\ion{Si}{IV}}
doublet does not exhibit the first and the last components that we see
in the {\ion{C}{IV}} (i.e., there are only three components), and
second, a single central component is not enough to model the central
part of the {\ion{Si}{IV}} features alone. More precisely, the central
component (number 3) must be splitted in two parts (3a and 3b) in
order to obtain a good fit for {\ion{Si}{IV}}; infact, such a feature
does not show saturation in its central part, while {\ion{C}{IV}}
does. This is clearly visible especially for {\ion{Si}{IV}
$\lambda$1402} \AA\ (Fig. 2). Thus, using four components to represent
{\ion{Si}{IV}}, the first and the last being fixed at the redhift of
the second and the fourth of {\ion{C}{IV}}, we obtain a good fit
($\chi^2 = 0.64$) for this feature.  All the further features we
analyzed have been fitted fixing the redshift of the components to
those we identified for either {\ion{C}{IV}} or {\ion{Si}{IV}}. Case
by case, we decided to fit groups of elements with either
{\ion{C}{IV}} or {\ion{Si}{IV}}, the choice depending on the extension
of the lines and the saturation of the central part of the
features. We specify that in these fitting procedures, the redshift of
the components has been linked to that of {\ion{C}{IV}} or
{\ion{Si}{IV}}, which in turn has been fixed to the values we find in
the {\ion{C}{IV}} and {\ion{Si}{IV}} fit. Similarly, the column
densities and Doppler parameters of the components of {\ion{C}{IV}}
and {\ion{Si}{IV}} has been fixed while fitting these pilot lines
together with other ones.

Finally, since the two central components of
{\ion{Si}{IV}} (3a and 3b) can not be resolved in the case of
{\ion{C}{IV}}, the corresponding column densities are computed
separately, but summed together in the analysis as if we were dealing
with a single component.

We detected several absorption fine structure features in the spectrum
at the GRB redshift. The gas can be excited to such states via
collisional effects (if the density is sufficiently high) or by the
absorption of UV photons followed by the spontaneous radiative decay
to an excited level of the ground state. Such lines can be very
important in determining some physical properties of the gas, and thus
we will consider them carefully. The next sub-subsections describe
the fits we performed to obtain the column densities of the elements
that show fine structure transitions. The column densities
corresponding to each component are reported in Table 2 and 3.

\subsubsection{The ion CII}

For this ion, we identified two fine structure doublets, namely
{\ion{C}{II} $\lambda$1036} \AA\ {\ion{C}{II*}
$\lambda$1037} \AA\ and {\ion{C}{II} $\lambda$1335} \AA\
{\ion{C}{II*} $\lambda$1337} \AA.  We fitted these doublets together 
with {\ion{C}{IV} } (Figs. 3 and 4). We note that only the second and 
third components out of the five in the fits in Figure 2 
are present in the excited fine structure levels, with a possible indication 
of the first component for {\ion{C}{II*} $\lambda$1037} \AA.
The fifth component, which is the one with the lowest redshift, is
present only for {\ion{C}{II} $\lambda$1036} \AA; the blending
of {\ion{C}{II*} $\lambda$1037} \AA\ by {\ion{C}{II}
$\lambda$1036} \AA.  makes impossible to say whether this component is
present in the excited level or not.  The {\ion{C}{II}
$\lambda$1335} \AA\ doublet appears to be saturated; this is evident
by looking at the best fit, which is completely flat in the center of
this line complex. Therefore, we use only the {\ion{C}{II}
$\lambda$1036} \AA\ doublet to evaluate the electron density and
temperature.

\begin{figure}
\centering
\includegraphics[angle=-90,width=9cm]{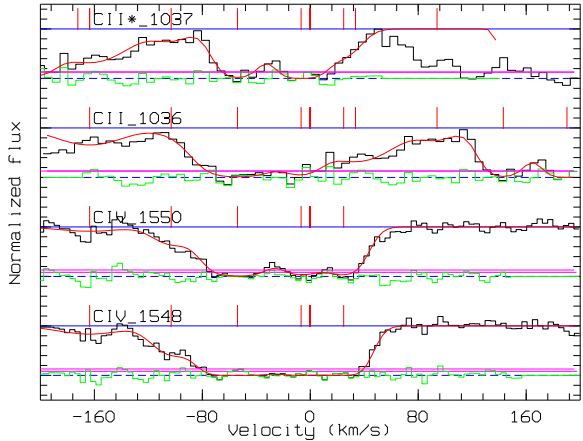}
\caption{The {\ion{C}{II} $\lambda$1036,1037*} \AA\ fine
structure doublet. The solid line represents the five Voigt components fit.  
}
\label{spe1}
\end{figure}

\begin{figure}
\centering
\includegraphics[angle=-90,width=9cm]{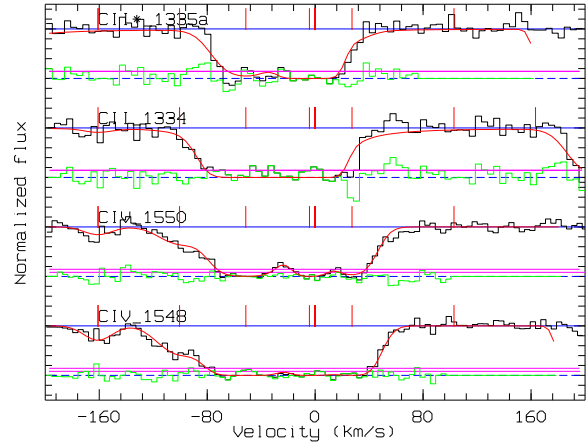}
\caption{The {\ion{C}{II} $\lambda$1334,1335*} \AA\ fine structure
doublet. The solid line represents the five Voigt components fit.
}
\label{spe1}
\end{figure}

\subsubsection{The ion SiII}

For this ion, we identified three fine structure doublets:
{\ion{Si}{II} $\lambda$1304} \AA\ ({\ion{Si}{II*}
$\lambda$1309} \AA), {\ion{Si}{II} $\lambda$1526} \AA\
({\ion{Si}{II*} $\lambda$1533} \AA) and {\ion{Si}{II}
$\lambda$1260} \AA\ ({\ion{Si}{II*} $\lambda$1265} \AA).  We
fitted the first two doublets together with {\ion{Si}{IV}}
(Figs. 5 and 6). The first of these doublets has been fitted with the
OI fine structure triplet too.

The third doublet falls in the Ly-$\alpha$ absorption gap. A reliable
analysis of this feature is difficult because a correct subtraction of
the continuum is not possible. Moreover, the excited level is further
splitted into two sublevels, which blend with each other and render it
impossible to fit the contribution of the various components.  

As for CII, only the second and possibly the third component is
present in the excited fine structure levels.  Finally, the third
component of the {\ion{Si}{II*} $\lambda$1533} \AA, if present,
would fall in a region rich of telluric features, and thus can hardly
be useful in determining the physical parameters of the gas.

\begin{figure}
\centering
\includegraphics[angle=-90,width=9cm]{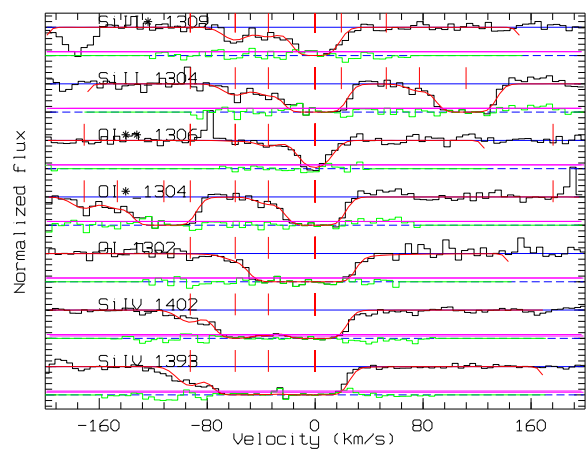}
\caption{The {\ion{Si}{II} $\lambda$1304,1309*} \AA\ fine structure 
doublet and {\ion{O}{I} $\lambda$1302,1304*,1306**} \AA\ fine structure
triplet.  The solid line is the four Voigt components fit.
}
\label{spe1}
\end{figure}

\begin{figure}
\centering
\includegraphics[angle=-90,width=9cm]{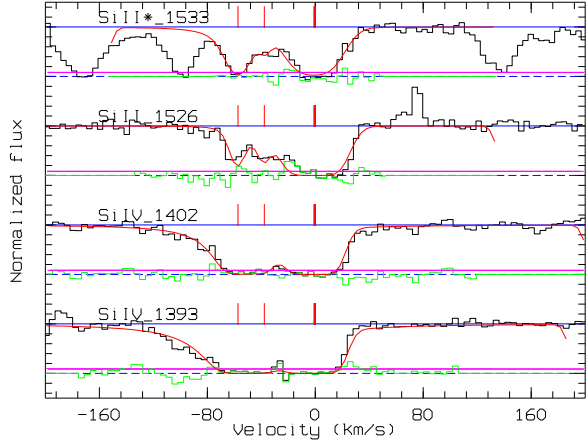}
\caption{The {\ion{Si}{II} $\lambda$1526,1533*} \AA\ fine
structure doublet. The solid line is the four Voigt components fit. The
$\lambda$1533 line is blended by several telluric features.
}
\label{spe1}
\end{figure}

\subsubsection{The atom OI}

This atom is present with the {\ion{O}{I}
$\lambda$1302,1304*,1306**} \AA\ fine structure triplet (Fig. 5). It has
been fitted with a four component profile, together with the
{\ion{Si}{IV} } components and the {\ion{Si}{II} 
$\lambda$1304,1309*} \AA\ fine structure doublet.  As for previous ions, 
the only components present in the excited fine structure levels are the 
second and the third (splitted in 3a and 3b). The second component 
appears to be saturated both in the ground and in the first excited level 
{\ion{O}{I} $\lambda$1304*} \AA, and will not be used in the analysis.

\subsubsection{The ion FeII}

This ion is present in a wide variety of flavors. In this 
paper, we concentrate on the fine structure multiplet around $\lambda =1600$
\AA. This multiplet shows the following absorption features:
{\ion{Fe}{II} $\lambda$1608} \AA, {\ion{Fe}{II}
$\lambda$1611} \AA, {\ion{Fe}{II*} $\lambda$1618} \AA,
{\ion{Fe}{II*} $\lambda$1621} \AA, {\ion{Fe}{II**}
$\lambda$1629} \AA, {\ion{Fe}{II***} $\lambda$1634} \AA,
{\ion{Fe}{II***} $\lambda$1636} \AA. Fig. 7 presents the
multiplet, together with the {\ion{Si}{IV} $\lambda$1393} \AA\
absorption feature.

We can see that the {\ion{Fe}{II}} excited levels only show 
the second component, while the ground state also includes the third
component. We obtain a good fit with a single Voigt profile for the
excited levels and a double line profile for the ground state. The
central wavelengths of these Voigt profiles have been fixed to the 
redshifts of the second and third components of {\ion{C}{IV}}.

\begin{figure}
\centering
\includegraphics[angle=-90,width=9cm]{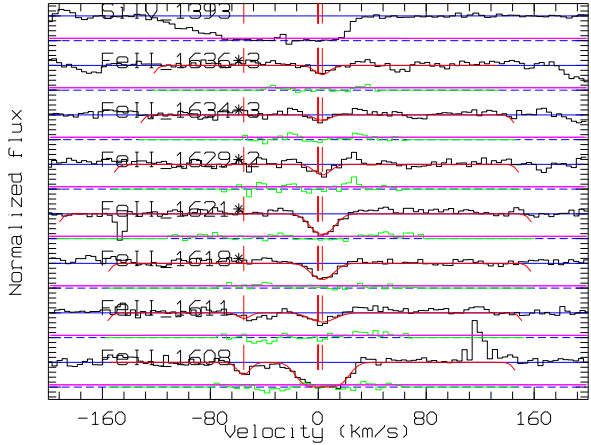}
\caption{The FeII 1608 fine structure multiplet, together with the SiIV 1393 absorption feature. The 
solid line represents the double Voigt component fit.
}
\label{spe1}
\end{figure}

\medskip 

In the next subsections we will describe in detail the physics which
relates the fine structure transitions to the density and temperature
of the gas that produces these features (provided that the fine levels
are excitated by collisional effects).

\subsection{Fine structure line excitation via collisions }

For ions with a doublet fine structure in the ground state, the ratio
between the population density of the higher and the lower level is:

$${ n_2\over n_1} = {{Q_{12}+w_{12}}\over{ Q_{21}+w_{21}}},\eqno (1)  $${}

\noindent
where $Q_{12}$ and $Q_{21}$ are the collision rates from level 1 to
level 2 and vice versa, and $w_{12}$ ($w_{21}$) is the photon
absorption (decay) rate between the two levels (Bahcall \& Wolf
1968). $Q_{12}$ and $Q_{21}$ can be written in the form $Q = \Sigma_j
<\sigma v>_jN_j$, where each kind of exciting particle (proton, 
electron, etc.) is denoted by $j$, $\sigma$ is the collisional cross
section for the process involving the $j$-th species, whose density and
average speed are $N_j$ and $<v>_j$, respectively.

We are interested in the situation in which level 2 is the excited
fine structure of the ground state 1. The excited fine structure
levels can be produced by photon pumping or collisional excitations.
Generally, the following mechanisms can be responsible for this
excitation: (i) direct excitation by IR photons; (ii) excitation to
upper levels through absorption of UV photons, followed by the
spontaneous decay to an excited level of the ground state (indirect UV
pumping); (iii) collisions with charged particles (mainly electrons)
or with neutral hydrogen. PCB06 showed that the excitation rate by UV
pumping is larger than IR pumping, rendering the latter excitation
mechanism unimportant for gas observed in GRB afterglow spectra.  In
the following, we will elaborate the scenario in which collisions are
the main mechanism responsible for the production of fine structure
transitions and discuss subsequently (section 4.4) the situation in
which indirect UV pumping is at work.

If we assume that the collisions between electrons and ions represent
the main process responsible for the excitation of the fine structure
levels, equation (1) simplifies since the quantities of the parameter
$Q$ now refer only to electrons. In particular, for the
{\ion{C}{II}} and {\ion{Si}{II}} fine structure doublets,
equation (1) becomes:

$$ n_e = {75\; T_4^{0.5}\over [2\; {N(CII)\over N(CII^*)}e^{91.75/T}-1]}, \eqno (2)$$

and

$$ n_e = {1.28\times 10^3\; T_4^{0.5}\over [2\; {N(SiII)\over N(SiII^*)}e^{413/T}-1]}, \eqno (3)$$

\noindent
respectively (Srianand \& Petitjean 2001). These equations have been
derived taking into account that the radiative excitation from the
ground to the excited state is forbidden, and using the statistical
weights; the radiative decay rates and the energy gaps are given in
Bahcall \& Wolf (1968).  Here $n_e$ is the electron density, $N$ is the
column density of the element in parentheses (the asterisk refers to
excited fine structure levels) and $T_4$ is the kinetic temperature in
units of $10^4\; K$.

For a fine structure system with multiple levels, the situation is
more complicated, unless we assume that electron-ion collisions are
dominant with respect to the radiative processes. If this is the case,
the $w$ terms in equation (1) can be neglected, and the electron
density disappears, yielding:

$$ {n_i \over n_j} = {g_i \over g_j}\; exp[-E_{ij}/kT], \eqno (4) $$

\noindent
in the case of thermal equilibrium (PCB06).  Equation (4) tells us
that the levels are populated according to a Boltzmann distribution;
here $g_i$ is the degeneracy of the state $i$, $E_{ij} = E_i-E_j$ is
the energy difference between the state $i$ and $j$ and $T$ is the
excitation temperature.

\begin{table*}[ht]
\caption{\bf Logarithmic ion column densities in cm$^{\bf -2}$ for the components of the main absorption system at z=3.967}
\footnotesize
\begin{tabular}{lcccccccc}
\hline
Velocity shift$^{a}$ (component)& {\ion{C}{II}} & {\ion{C}{IV}} & {\ion{O}{I}} & {\ion{Si}{II}} & {\ion{Si}{IV}} & {\ion{Fe}{II}} \\
\hline
+32.6  (1)   & Blend  &  14.4 $\pm$ 0.08 & NO  & NO & NO & NO \\
+2.4   (2)   & 14.33 $\pm$ 0.15 &  14.86 $\pm$ 0.08 & SAT            & 14.68 $\pm$ 0.23 & 15.53 $\pm$ 0.11 & 15.24 $\pm$ 0.06 \\
-44.0  (3a)   & 14.81 $\pm$ 0.13 &  14.75 $\pm$ 0.12 & 14.94 $\pm$ 0.09 & 13.62 $\pm$ 0.06 & 13.72 $\pm$ 0.09 & 14.66 $\pm$ 0.22 \\
-44.0b (3b)   &                &                 & 13.43 $\pm$ 0.11 & 13.60 $\pm$ 0.07 & 14.11 $\pm$ 0.10 &                  \\ 
-90.2  (4)  & 12.98 $\pm$ 0.3  &  13.57 $\pm$ 0.13 & NO             & 12.53 $\pm$ 0.24 & 13.21 $\pm$ 0.03 & NO               \\
-154.6 (5)  & 12.75 $\pm$ 0.35 &  13.04 $\pm$ 0.06 & NO            .&    NO          & NO             & NO               \\
\hline
\end{tabular}
\normalsize

$^{a}$km/s

\end{table*}

\begin{table*}[ht]
\caption{\bf Logarithmic ion column densities of excited fine structure states in cm$^{\bf -2}$ for the components of the main absorption 
system at z=3.967}
\footnotesize
\begin{tabular}{lcccccccc}
\hline
Vel.$^{a}$ (Comp)& {\ion{C}{II*}} & {\ion{O}{I*}} & {\ion{O}{1**}} & {\ion{Si}{II*}} & {\ion{Fe}{II*}} & {\ion{Fe}{II**}} & {\ion{Fe}{II***}}\\ 
\hline
+2.4 (2)                & 14.69 $\pm$ 0.26 & SAT               & 14.64 $\pm$ 0.05 & 14.44 $\pm$ 0.09 & 14.29 $\pm$ 0.03   & 13.74 $\pm$ 0.08 & 13.60 $\pm$ 0.09 \\
-44.0 (3a)               & 14.55 $\pm$ 0.09 & 13.56 $\pm$ 0.08   & 13.46 $\pm$ 0.09 & 13.41 $\pm$ 0.15 & $< 13.14$            & NO               & NO      \\
-44.0b (3b)              &                  &  NO               & 12.55 $\pm$ 0.60 & 13.66 $\pm$ 0.11 & NO                 & NO               & NO      \\
\hline
\end{tabular}
\normalsize

$^{a}$km/s
\end{table*}

\begin{figure}
\centering
\includegraphics[angle=0,width=9cm]{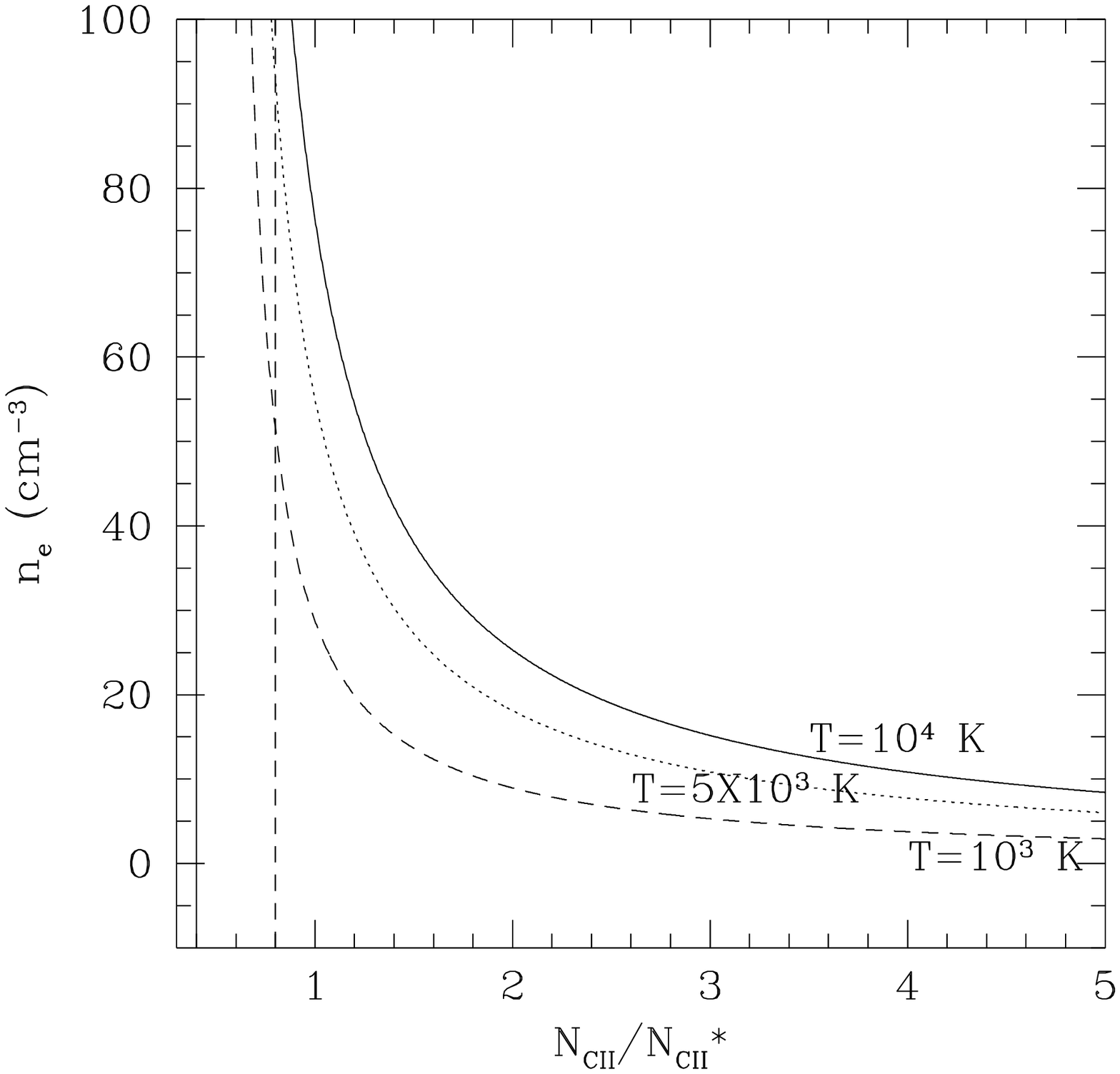}
\includegraphics[angle=0,width=9cm]{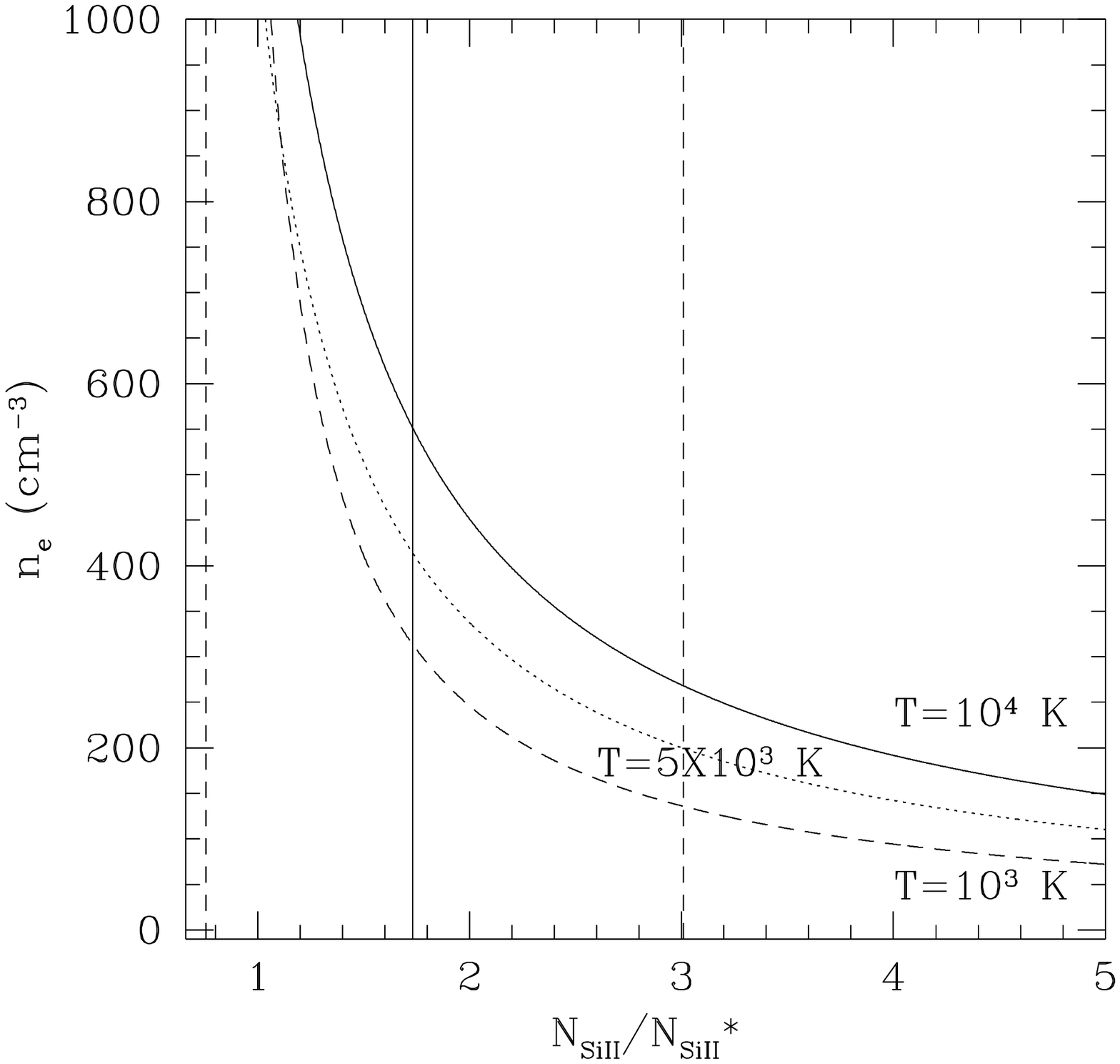}
\caption{Top panel: the electron density as a function of the ratio between the 
absorbing columns of the ground state with respect to the excited one, 
for three different values of the temperature, calculated for the 
{\ion{C}{II}} fine structure doublets. Vertical lines represent the 
observed ($N/N^*$) ratio (solid line) and its $1 \sigma$ errors (dashed 
lines). The figure refers to the second component of the host 
environment gas. Bottom panel: same as top panel, but for {\ion{Si}{II}} 
fine structure doublets.
}
\label{spe1}
\end{figure}

\subsection{Temperature and density of the gas in the case of collisional excitation}

In this section we try to obtain an estimate of the temperature and
density of the gas in the GRB environment. We showed in the previous
section that the excited states of the fine structure transitions 
present some absorption only in the second and third component of the
five identified.  For this reason, the physical quantities can be
computed or constrained only in these two cases.

We proceed in the following way. First of all, we assume 
that the electron-ion collisions are the main mechanism 
responsible for the production of the excited fine structure feature (PCB06).

For fine structure doublets ({\ion{C}{II}} and
{\ion{Si}{II}}), we plot in Fig. 8 the electron density as a
function of the ratio between the absorbing columns of the ground
state with respect to the excited one ($N/N^*$), using eqs. (2) and
(3), and for different values of the temperature. In this plot,
vertical lines are loci of constant $N/N^*$.  We then draw vertical
lines corresponding to our measure of the ratio $N/N^*$ and to the
corresponding errors. In this way we can constrain the density of the
gas for a fixed interval of temperature, and even constrain the
temperature itself, if both the {\ion{C}{II}} and
{\ion{Si}{II}} present an excited fine structure absorption.

For fine structure multiplets, we use eq. 4 under the assumpion of
thermal equilibrium. Fitting the ratio of the column densities to
their multiplicity factors with respect to the energy difference
between each excited state and the ground state, we can obtain a
measure of the temperature of the gas. Such a temperature can
constrain the gas density using the plots of eqs. (2) and (3) for the
fine structure doublets, or can be used as an input for the PopRatio
software package (Silva \& Viegas 2001), to estimate the electron
density independently once the ratio between the column densities of
the fine structure levels of the multiplets is known.

As previously remarked, only component 2 and 3 of the main absorption
system feature fine structure transition; in the following
sub-subsections we present the analysis of such components.

\subsubsection{Component z=3.967\_2}

This component features high column densities for all the ground
states: {\ion{C}{II}}, {\ion{O}{I}}, {\ion{Si}{II}}
and {\ion{Fe}{II}}.  The excited fine structure levels appear in
all their flavors: we have strong {\ion{C}{II*}},
{\ion{O}{I*}} / {\ion{O}{I**}} and {\ion{Si}{II*}}
columns; moreover {\ion{Fe}{II}} is present with its first
three excited levels.  We plot the electron density as a function of
($N/N^*$) for three different temperatures, namely $T=10^4$,
$T=5 \times 10^3$ and $T=10^3$ K, for {\ion{C}{II}} (Fig. 8,
left panel) and {\ion{Si}{II}} (Fig. 8, right panel). For
{\ion{C}{II}} we used the 1036-1037 \AA\ feature, because
{\ion{C}{II} $\lambda$1334} \AA\ is saturated.

We can see from Fig. 8 that both for {\ion{C}{II}} and
{\ion{Si}{II}} the observed column densities can only put a
lower limit to the electron density, independently of the temperature of the 
gas. This lower limit lies in the range $50 \div 200$ cm$^{-3} $ and 
$100 \div 300$ cm$^{-3}$ for temperatures of 
$10^3 \div 10^4$ K, for {\ion{C}{II}} and  
{\ion{Si}{II}} respectively.

For this component, we also made a fit to the temperature using eq. (4) and
the {\ion{Fe}{II}} multiplet data (Fig. 9). The value we found is
$T=2600^{+3000}_{-900} $ K (90\% confidence limit); such a value agrees
with that found by PCB06, for the same GRB.

The same fit has not been performed for the {\ion{O}{I}}, because 
its ground state appears to be saturated.

\begin{figure}
\centering
\includegraphics[angle=-90,width=9cm]{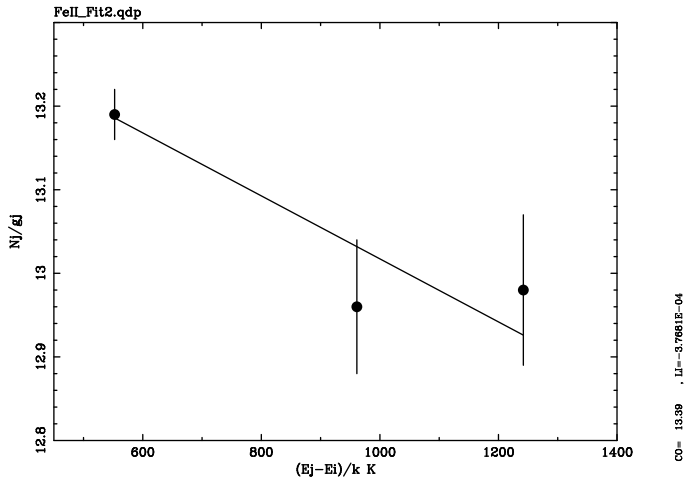}
\caption{The ratios between column density and degeneracy of the excited fine 
structure {\ion{Fe}{II}} levels as a function of their energy gap 
to the ground state. The solid line represents our best fit to the 
data.
}
\label{spe1}
\end{figure}

\begin{figure}
\centering
\includegraphics[angle=0,width=9cm]{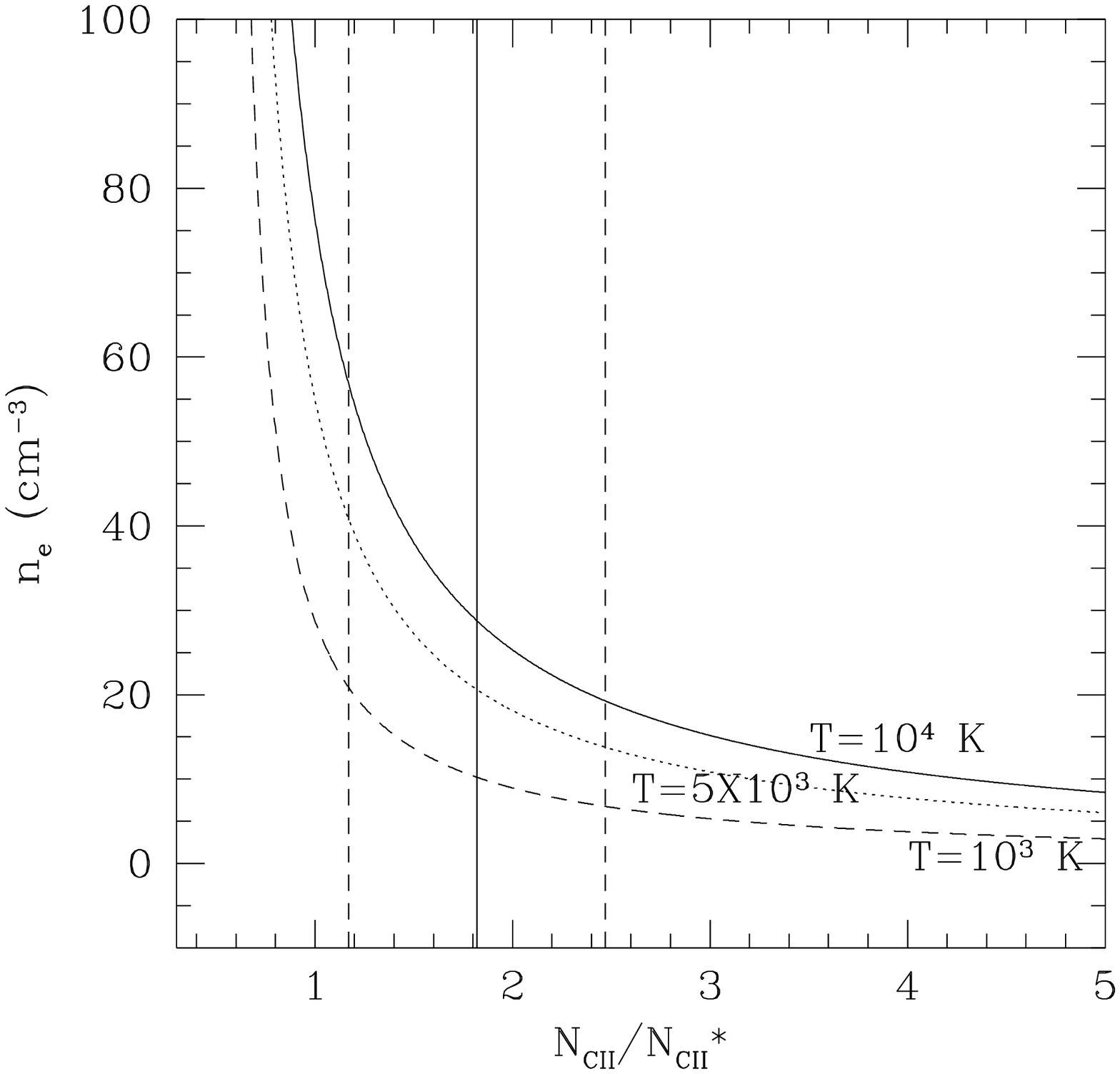}
\includegraphics[angle=0,width=9cm]{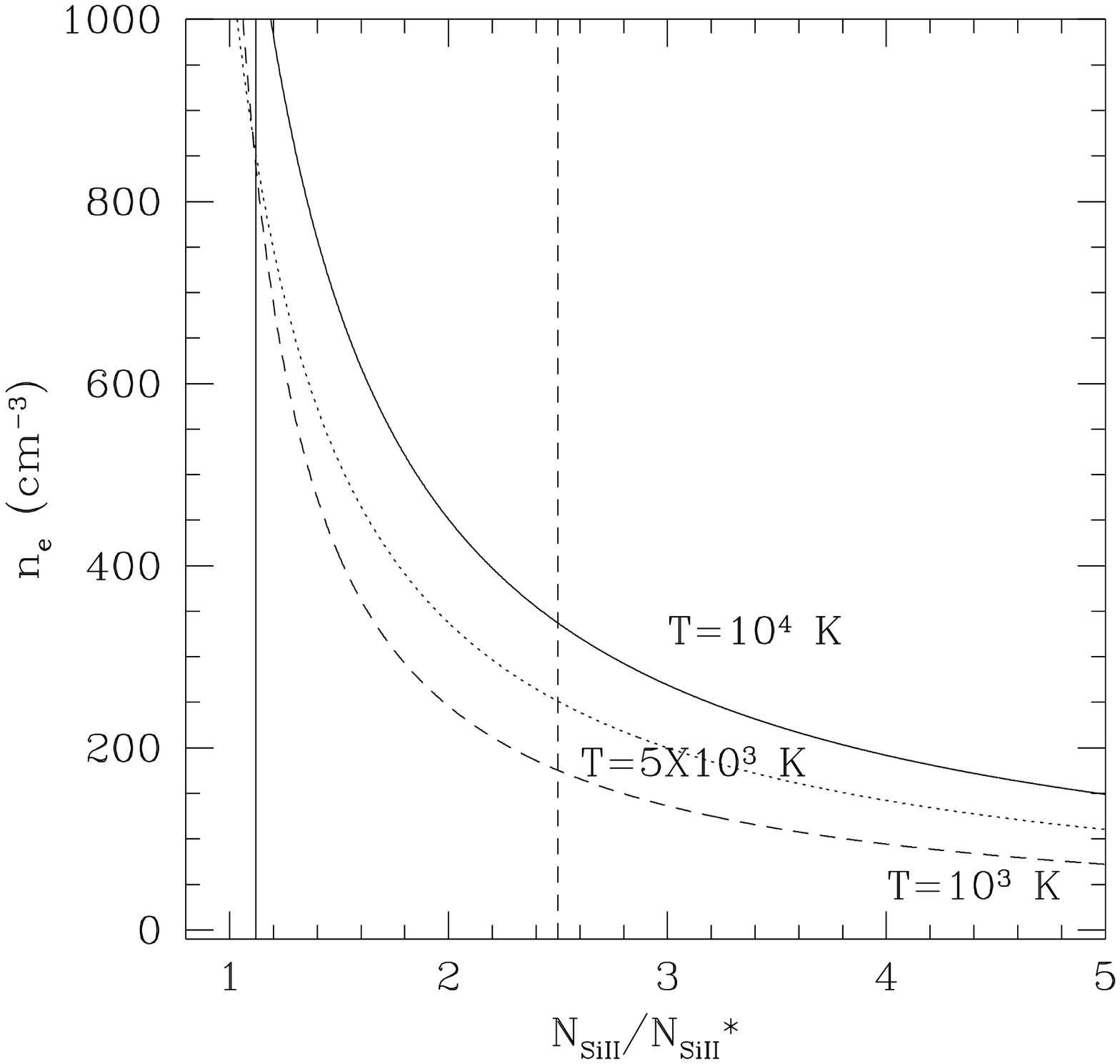}
\caption{Top panel: the electron density as a function of the ratio between the 
absorbing columns of the ground state with respect to the excited one, 
for three different values of the temperature, calculated for the 
{\ion{C}{II}} fine structure doublets. Vertical lines represent the 
observed ($N/N^*$) ratio (solid line) and its $1 \sigma$ errors (dashed 
lines). The figure refers to the third component of the host environment 
gas. Bottom panel: same as top panel, but for {\ion{Si}{II}} fine 
structure doublets.  
}
\label{spe1}
\end{figure}

\subsubsection{Component z=3.967\_3}

This component features high column densities for the ground states of
the {\ion{C}{II}}, {\ion{O}{I}} and {\ion{Fe}{II}},
while that of {\ion{Si}{II}} is very low. As for the excited
fine structure levels, we have strong {\ion{C}{II*}} and low
{\ion{O}{I}} columns; {\ion{Fe}{II}} excited levels are
absent.  {\ion{Si}{II*}} is difficult to treat. Apart from the
$\lambda$1265 line that falls in the Ly-$\alpha$ absorption gap, the
$\lambda$1533 line is blended with several telluric features
(Fig. 6). The last flavor of {\ion{Si}{II}}, the $\lambda$1309
line, shows a hint of a possible absorption feature (Fig. 5); however,
since the corresponding ground state is very lowly populated, this feature is
somewhat suspect. As for the second component, we plot the electron
density as a function of ($N/N^*$) for three different temperatures,
namely $T=10^4$, $T=5 \times 10^3$ and $T=10^3$ K, for
{\ion{C}{II}} and {\ion{Si}{II}} (Fig. 10). For
{\ion{C}{II}} we used the $\lambda$ 1036-1037 feature, because
{\ion{C}{II} $\lambda$1334} \AA\ is saturated.

Fig. 10, left panel indicates that the electron density inferred from
{\ion{C}{II}} lies in the range $10 \div 60$ cm$^{-3}$ for a
temperature of $10^3 \div 10^4$ K. If we consider the plot for
{\ion{Si}{II}}, the density appears to be higher than $\sim 200$
cm$^{-3}$, if we assume a temperature of $\sim 10^3$ K. Thus, there is
no overlap between the two solutions. As we discussed above, the
{\ion{Si}{II*} $\lambda$1309} \AA\ is a suspicious feature, because
the corresponding ground state ({\ion{Si}{II} $\lambda$1304} \AA) is
poorly populated, while for the other elements the ground state is
very highly populated when an excited level is present. In the
following, we will consider either that {\ion{Si}{II*} $\lambda$1309}
\AA\ is real or it is not.

For this component, we also estimate a temperature using eq. 4 and
 {\ion{O}{I}} triplet data. The value we find is $T=10^3$ K, 
but we only  have two points in the fit so the errors are too 
large to  put an acceptable constraint on the temperature.

\subsection{Fine structure line excitation via indirect UV pumping}

In the previous sections we assumed that fine structure excited levels
are produced by collisions. The other, competitive mechanism is
indirect UV pumping. In fact, the fine structure levels can be excited
by the absorption of a UV photon to an upper level followed by the
spontaneous decay to an excited lower level. If this mechanism is at
work, we cannot use the fine structure abundances to gather
information about the temperature and density of the gas, but only on
the strength of the radiation field. PCB06 (using the PopRatio code
developed by Silva \& Viegas 2001, 2002) analyzed the relation
between the far UV radiation field intensity and the relative fraction
of excited fine structure states with respect to their ground levels
(for {\ion{O}{I}}, {\ion{Si}{II}} and
{\ion{Fe}{II}}, see their Figs. 7 and 8). In the following
subsections, we will use their plots to estimate the radiation field
intensity from our column density data, in the case that the fine
structure transitions are produced by indirect UV pumping.

\subsubsection{Component z=3.967\_2}

As we showed before, this is the component which presents more fine structure 
excited states. We evaluate the ratio between the ground and the 
first excited levels for {\ion{Fe}{II}}($\lambda$1608-1611 \AA
and $\lambda$1618*-1621* \AA) as well as the ratio between the 
{\ion{Si}{II} $\lambda$1526} \AA\ and {\ion{Si}{II*} $\lambda$1533} 
\AA\ fine structure levels. The {\ion{O}{I}} ground state is 
saturated and can not be used in the analysis. Using Fig. 7 of PCB06, 
we find that the two ratios are compatible with a radiation field 
intensity of $G/G_0 = 5\times 10^5$, $G_0 = 1.6\times 10^{-3}$ erg cm$^{-2}$
s$^{-1}$ being the Habing constant. A consistent result is obtained 
estimating the ratios of the absorbing column of the first excited level 
of FeII to the second and third ones, and using Fig. 8 of PCB06. Again, 
we find that the two ratios yield a value of $G/G_0 = 
5\times 10^5$. The 90\% confidence interval for the field intensity is 
$G/G_0=10^5 \div 10^6$.

\subsubsection{Component z=3.967\_3}

This component exhibits the {\ion{C}{II}} fine structure doublet, the
{\ion{O}{I}} triplet and possibly the {\ion{Si}{II}} excited level.
As we discussed in section 4.3.2, the {\ion{Si}{II*}} detection is
uncertain.  Nevertheless, using the {\ion{O}{I}} and {\ion{Si}{II}}
ratios between the ground and first excited level, we find two
different values for the radiation field intensity: $G/G_0 = 10^5$ and
$G/G_0 = 10^6$, respectively. These values do not overlap at the 90\%
confidence limit. Since the detection of {\ion{Si}{II*}} is uncertain,
we cannot exclude in this way that the fine structure excited levels
in this component are produced by indirect UV pumping. However, the
theoretical lines representing the {\ion{Fe}{II*}}/{\ion{Fe}{II}} and
{\ion{O}{I*}}/{\ion{O}{I}} abundance ratios as a function of the
radiation field in Fig. 8 of PCB06, approach each other very
closely. This means that we should observe a similar column of
{\ion{Fe}{II*}} with respect to {\ion{Fe}{II}}, as we observe for
{\ion{O}{I*}} with respect to {\ion{O}{I}}. Since $N_{OI*}/N_{OI} =
0.04$, and $N_{FeII} = 14.66$ cm$^{-2}$ (see Tab. 2), we expect
$N_{FeII}* = 13.26$ cm$^{-2}$.  However, this is not the case, because
the upper limit that we found for the first excited level of the
{\ion{Fe}{II}} is $N_{FeII}* < 13.14$ cm$^{-2}$ at the 90\% confidence
level. On the other hand, if the fine structure states of the third
component are produced by collisions, the {\ion{O}{I*}}/{\ion{O}{I}}
abundance ratio would yield an electron density of a few tens cm$^{-3}$,
compatible with that found for the {\ion{C}{II}} fine structure
doublet in section 4.3.2, and the ratio {\ion{Fe}{II*}}/{\ion{Fe}{II}}
would be $10$ times smaller the {\ion{O}{I*}}/{\ion{O}{I}} one (See
Fig 9 in PCB06). This means that we expect $N_{FeII}* = 11.96$
cm$^{-2}$, a value consistent with the observed upper limit.

\subsection{Neutral elements and gas distance from the GRB}

The presence of neutral elements in the spectra of GRB afterglows is
important, because it can put strong constraints on the distance of
the gas from the GRB. This is because the first ionization potential
of the elements usually lies below the 1 Ryd threshold. Thus, the
intense radiation coming from the GRB would easily ionize the neutral
elements, without the screening of the Hydrogen atoms, the first
ionization potential of which is at higher energies. PCB06 observed 
the {\ion{Mg}{I} $\lambda$2852} \AA\ transition in GRB051111, and
they put a lower limit of $50$ pc to the distance of the gas from the
GRB. Moreover, they refined this limit to $80$ pc since no variability of
the {\ion{Mg}{I}} line has been observed.

In GRB050730, we observe many {\ion{O}{I}} transitions.
Oxygen, however, has a first ionization potential that lies just above
the 1 Ryd threshold, and can exist in its neutral state even in the
presence of a strong radiation field, because the Hydrogen screens it
from the UV photons. We can not observe the {\ion{Mg}{I}
$\lambda$2852} \AA\ transition because of the redshift of our GRB. In
addition, we do not observe any other neutral element in the
afterglow so we can not put any constraint on the gas distance from
the GRB this way.  The upper limit that we find for the neutral carbon
absorption line ({\ion{C}{I} $\lambda$1656} \AA) is $N_{CI}* <
13.2$ cm$^{-2}$ at the 90\% confidence level.

Some considerations on the relative distance of the different
shells of gas from the GRB can be pursued. Component 1 of the z =
3.967 system, the one with the longest wavelength, highest positive
velocity shift (see sec. 4.1), does not present any fine structure or
low ionization lines; it only shows very high ionization features,
such as {\ion{C}{IV}} and {\ion{O}{VI}}, suggesting that this
component is very close to the GRB site. Moreover, if we assume that
the fine structure lines of the shells are produced by the same
mechanism (collisions or UV pumping), we can also say something about
component 2 and 3. In fact, in the case of collisional excitations
we, can estimate the electron density of the gas and this can be
related to the distance $r$ using the definition of the ionization
parameter:

$$ U = {L_{phot}\over 4 \pi r^2  c n_e}, \eqno (5)$$

\noindent
where $L_{phot}$ is the number of ionizing photons and $c$ is the
speed of light. A time independent estimate of $U$ with
photoionization codes like CLOUDY (Ferland 2002) is inappropriate in
the case of a strongly variable phenomenon such as a
GRB. Nevertheless, if we assume the electron density found in section
4.3 for the third component ($10 \div 100$ cm$^{-3}$) and for the
second one ($10^3 \div 10^6$ cm$^{-3}$) we can conclude (using the
scaling law of eq. 5) that the gas belonging to the third component
lies $10 \div 100$ times farther from the GRB than that belonging to
the second one.

In a similar way, we can assume that indirect UV pumping is at
work. In this case, the distance increases as the square root of the
UV flux. If we trust the estimate of the radiation field
intensity evaluated only from the
{\ion{O}{I*}}/{\ion{O}{I}} ratio for the third
component, we can conclude that the gas in the second component is a
few ten times closer to the GRB than that of the third one.

\subsection{Metallicity}
                                                                                    
\begin{figure}
\centering
\includegraphics[angle=-90,width=11cm]{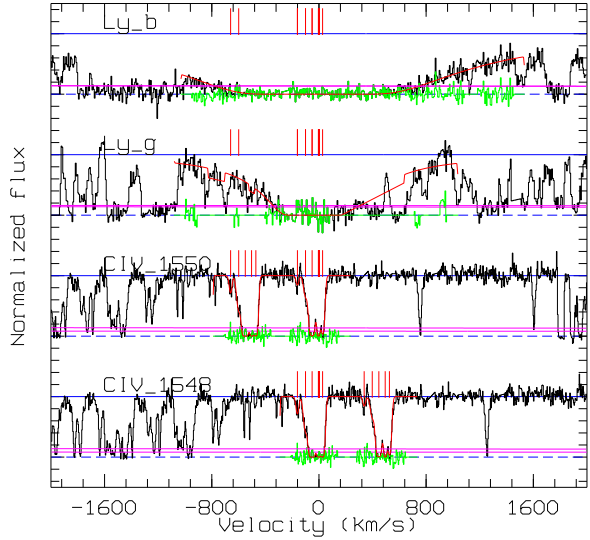}
\caption{The Ly-$\beta$ and Ly-$\gamma$ absorption features, together with
the {\ion{C}{IV} $\lambda$1548} and {\ion{C}{IV}
$\lambda$1550} doublet. The red line represents the five Voigt
component fit.
}
\label{spe1}
\end{figure}

Fig. 11 shows the Ly-$\beta$ and Ly-$\gamma$ absorption
features. It is impossible to discriminate between the five components
identified from the analysis of {\ion{C}{IV}} and
{\ion{Si}{IV}}. The hydrogen absorption features result in very
broad lines with two extended wings, which reach well beyond the
velocity range of the heavier ions. Since a component
by component analysis is not possible, we can only estimate the
metallicity of the circumburst material in terms of the ratio [X/H] between the
total absorption column of the X element with respect to that of
hydrogen. Since the latter  mainly includes hydrogen
in the outer regions of the galaxy, we will underestimate the
metallicity. 

\begin{table*}[ht]
\caption{\bf Metallicity}
\footnotesize
\begin{tabular}{lcccc}
\hline
Ion  & $\log N$ (cm$^{-2}$) & $\log N^{(a)}$ (cm$^{-2}$) & [X/H] & [X/H]$\sun$ \\
\hline
{\ion{H}{I}} & $22.05 \pm 0.29$ &  -                  &  -
     &  -                   \\
{\ion{C}{II}} & $15.23 \pm 0.15$ &                     &
     &                      \\
{\ion{C}{IV}} & $15.19 \pm 0.09$ &                     &
     &                      \\
C                   &                  & $15.51 \pm 0.12$    &  $-6.54 \pm 0.41$
&  $-2.93 \pm 0.41$    \\
{\ion{O}{I}} & SAT              &                     &
     &                      \\
{\ion{O}{VI}} & SAT              &  -                  &  -
     &  -                   \\
{\ion{Si}{II}} & $14.96 \pm 0.15$ &                     &
     &                      \\
{\ion{Si}{IV}} & $15.55 \pm 0.12$ &                     &
     &                      \\
Si                   &                  & $15.65 \pm 0.13$    &  $-6.40 \pm 0.42$
&  $-1.94 \pm 0.42$    \\
{\ion{Fe}{II}} & $15.39 \pm 0.10$ & $15.39 \pm 0.10$    &  $-6.66 \pm 0.39$
&  $-2.11 \pm 0.43$    \\
\hline
\end{tabular}
\normalsize                                                                 
$^{(a)}$ Total column density of the element
\end{table*}

Table 4 shows our estimate of the metallicity. The second column
presents the column densities of the different ions, being the sum of
the columns of the five components of the ground states and of the
excited fine structure levels, if these are present. The third column
of Table 4 is the total column density for each element, and is
obtained adding the column densities of the various ions through which
the element appears in the spectrum. We could not get an estimate of
the column density of Oxygen, because the absorption features of
{\ion{O}{I}} and {\ion{O}{VI}} are saturated.  Not surprisingly, we
obtain a metallicity of $10^{-2} \div 10^{-3}$ relative to the solar
one, in agreement with the paper by Starling et al. (2005). It should
also be noted that these estimates are not corrected for depletion in
dust grains. In the Galactic ISM only $\sim1\%$ of the iron is in the
gas phase while the rest is locked in dust grains (Savage \& Sembach
1996). The situation can be very different in a GRB host galaxy, and
in particular, in the GRB surrounding medium (Waxman \& Draine 2000,
Draine \& Hao 2002, Perna \& Lazzati 2002, Perna, Lazzati \& Fiore,
2003), because the intense GRB radiation field can efficiently destroy
dust grains. Caution should therefore be used in using mean
metallicity estimates.
                                                                                    
We can also calculate the ratio [C/Fe], representative of the
enrichment of the $\alpha$ elements relative to iron. We measure a
mean [C/Fe]=0.08$\pm$0.24, consistent with the value predicted by the
models of Pipino \& Matteucci (2004, 2006) for a galaxy younger than 1
Gyr subject to a burst of star-formation.  In such a case Pipino \&
Matteucci (2006) also predict a low [Fe/H] value, close to that in
Table 4.  Again, we warn that mean abundance estimates may be affected
by large systematic uncertainties.

A safer approach is to estimate the [C/Fe] of each component of the
z=3.967 system.  We find that [C/Fe] of component 2 is -0.15$\pm$0.13
while that of component 3 is +0.53$\pm$0.23.  Indeed, note in Tables 2
and 3 that while the total carbon column density of component 2 and 3
are similar, the total colum density of iron of component 3 is about
four times less than that of component 2. A similar conclusion applies
to silicon.  The total silicon column density of component 3 is also
about 10 times less than that of component 2.  Interestingly, Perna,
Lazzati \& Fiore (2003) found that silicates tend to be destroyed more
efficiently than graphite if a dusty medium is exposed to the intense
GRB radiation field. This would leave more iron free in the gas phase
in clouds closer to the GRB site than in farther clouds, in agreement
with our suggestion that component 2 is 10-100 times closer to the GRB
site than component 3.

\section{Conclusions}

In this paper we present high resolution (R=20000-45000, corresponding
to 14 km/s at 4200\AA\ and 6.6 km/s at 9000\AA) spectroscopy of the
optical afterglow of Gamma Ray Burst GRB050730, observed by UVES@VLT
$\sim 4$ hours after the trigger.
                                                                                    
We confirm that the redshift of the host galaxy is z=3.96764. Four
intervening systems between z = 3.56 and z = 1.77 have been identified
along the GRB line of sight, in addition to the main system at the
redshift of the host galaxy.
                                                                                    
For what concerns the main absorption system, the spectrum shows
that the ISM of the GRB host galaxy is complex, with at least five
components contributing to this main absorption system at z=
3.967. Such components are identified in this paper with progressive
numbers with decreasing velocity values.

These absorption lines appear in different flavors, from neutral
hydrogen (Ly-$\alpha$, Ly-$\beta$, Ly-$\gamma$), to neutral
metal-absorption lines ({\ion{O}{I}}), and via low ionization
lines ({\ion{C}{II}}, {\ion{Si}{II}},
{\ion{Mg}{II}}, {\ion{Al}{II}}, {\ion{Ni}{II}},
{\ion{Fe}{II}}, {\ion{P}{II}}, {\ion{S}{II}}) to
high-ionization absorption features ({\ion{C}{IV}},
{\ion{Si}{IV}}, {\ion{N}{V}}, {\ion{O}{VI}}).
                                                                                    
In addition, we detect strong {\ion{C}{II*}},
{\ion{Si}{II*}}, {\ion{O}{I*}} and {\ion{Fe}{II*}}
fine structure absorption features in components 2 and 3 (no
{\ion{Fe}{II*}} in the last component). The ratio of these lines
to their ground states are used to constrain the gas density and
temperature, under the assumption that they are produced via
collisional excitation, or alternatively by UV pumping.  If
collisional excitation is at work, we find that the gas producing the
fine structure features has a temperature of a few thousand K, and an
electron density $n_e =10^3\div 10^6$ cm$^{-3}$ (component 2) or $n_e
=10\div 10^2$ cm$^{-3}$ (component 3). On the other hand, if UV
pumping is the process responsible for the fine structure features,
the radiation field exciting the gas has an intensity of $G/G_0 = 10^5
\div 10^6$.
                                                                                    
We speculate that the excitation mechanisms of the fine structure
lines may be different, with component 3 probably excited by
collisions, and component 2 by UV pumping. This is because the
observed {\ion{O}{I*}}/{\ion{O}{I}} ratio in component 3
would imply a similar {\ion{Fe}{II*}}/{\ion{Fe}{II}} ratio
in the case of UV pumping, which is not observed. On the other hand,
the radiation field intensities calculated from the fine structure
ratios for component 2 are all in agreement with each other.
                                                                                    
Component 1, the one with the longest vawelenght, highest
positive velocity shift, does not present any fine structure or low
ionization lines; it only shows very high ionization features, such as
{\ion{C}{IV}} and {\ion{O}{VI}}, suggesting that this component is
very close to the GRB site.

For the first three components we can derive information on the
relative distance to the site of the GRB explosion.  Both for
collisional excitation and UV pumping, component 2 appears to be 10-100
times closer to the GRB site than component 3. 
                                                                                    
As to the metallicity, for the system at z=3.967 we obtain mean values
$\approx 10^{-2}$  of the solar metallicity or less. However, this
should not be taken as representative of the circumburst medium,
since the main contribution to the hydrogen column density comes from
the outer regions of the galaxy while that of the other elements
presumably comes from the ISM closer to the GRB site and since the
correction for dust depletion is highly uncertain.  It is however
interesting to note that the mean [C/Fe] ratio agrees well with that
expected by the single star-formation event models of Pipino and
Matteucci (2004, 2006).

We could also measure the [C/Fe] of the single components 2 and 3. The
[C/Fe] ratio of component 2 is -0.15$\pm$0.13, marginally different
from that of component 3, 0.54$\pm$0.23. Although the result is
significant at only 2.6$\sigma$ confidence level, it is interesting to
speculate on its origin. If the cloud responsible for the absorption component 2
is closer than the cloud producing component 3 to the GRB site, we expect
that dust destruction in this cloud, due to the interaction of the GRB
radiation field with the medium, is more efficient than in the cloud
responsible for component 3.  Since silicates tend to be destroyed
more efficiently than graphite (Perna, Lazzati \& Fiore 2003), we
expect more iron and silicon freed in the gas phase in cloud 2 than in
cloud 3, as observed.

\begin{acknowledgements}
We acknowledge support from contract ASI/I/R/039/04 (Swift). FF
acknowledges support from contract ASI/I/R/023/05/0.  We thank Roberto
Maiolino for useful discussions. RP and DL ackowledge support by NSF
grant AST-0507571

\end{acknowledgements}

\end{document}